# The Explore of Knowledge Management Dynamic Capabilities, AI-Driven Knowledge Sharing, Knowledge-Based Organizational Support, and Organizational Learning on Job Performance: Evidence from Chinese Technological Companies.

Jun Cui[1,a,*]

[1] Solbridge International School of Business, Woosong University, Ph.D., Daejeon, Republic of Korea
[a] jcui228@student.solbridge.ac.kr, [b] email
*Corresponding author; Jun Cui (Email: jcui228@student.solbridge.ac.kr)

***Abstract:*** *Drawing upon Resource-Based Theory (RBT) and the Knowledge-Based View (KBV), this study investigates the impact of Knowledge-Based Organizational Support (KOS), AI-Driven Knowledge Sharing (KS), Organizational Learning (OL), and Knowledge Management Dynamic Capabilities (KMDC) on Organizational Performance (OP) in Chinese firms. In particular, this research explores the relationships among these factors, alongside control variables such as education level, staff skills, and technological innovation, to provide a comprehensive understanding of their influence on performance management. While recent studies on organizational performance have predominantly concentrated on digital business strategies and high-level decision-making, limited attention has been given to the role of digital maturity, workplace activities, and communication-related dynamics. This study addresses these gaps by consolidating critical factors that contribute to overarching job performance within organizations. Moreover, to empirically test the proposed hypotheses, data were collected from 129 valid questionnaires completed by employees across various Chinese firms. The research employed confirmatory factor analysis (CFA) to validate the measurement constructs and structural equation modeling (SEM) to evaluate the hypothesized relationships. The findings reveal several significant insights: (1) KOS, KS with AI, KMDC, and OL each have a direct positive effect on OP, emphasizing their critical roles in enhancing organizational outcomes. (2) Control variables, including education level, staff skills, and technological innovation, significantly moderate the relationships between KOS, KS with AI, KMDC, OL, and OP, further amplifying their impact. These results highlight the importance of fostering collaborative knowledge innovation mechanisms and leveraging dynamic capabilities for effective performance management. In summary, By integrating theoretical perspectives with robust empirical analysis, this study contributes to the literature on knowledge management and organizational performance, offering actionable insights for practitioners aiming to optimize performance in digitally transformative environments.*

## 1. Introduction

Over the past decade, some technological firms in many industries have seen the rapid development of Knowledge Management, AI technologies, Knowledge Technologies, Knowledge Sharing, and Organizational Learning. Moreover, the study indicates that some knowledge factors, AI technologies and the firms' organizational learning factors for the study by providing Chinese technical firm's background information. Similarly, knowledge management, knowledge sharing, and continuous learning and training have emerged as critical developmental priorities for numerous technological companies in China (Chowdhury et al., 2022). This study specifically focuses on Chinese software companies, with a prominent Chinese network security enterprise serving as the research subject and target. Employing quantitative analysis methods, the study incorporates reliability analysis and Partial Least Squares Structural Equation Modeling (PLS-SEM) to examine the proposed relationships and hypotheses. To facilitate robust data collection, electronic questionnaires were utilized, complemented by the use of widely adopted social platforms such as WeChat, WhatsApp, and LinkedIn. These tools enabled the efficient gathering of sample data, ensuring broad participation and representativeness within the study population. This methodological approach underscores the study's rigor in leveraging modern data collection and analysis techniques to explore the dynamics of knowledge management and organizational performance in the context of Chinese technological firms.

Despite the relatively modest sample size of 129, this study highlights the transformative role of AI and artificial intelligence technologies in shaping the core practices of knowledge management, sharing, and continuous learning within enterprises. These advancements significantly enhance organizational performance, productivity, revenue, and profitability, thereby driving efficiency improvements in Chinese technological companies. This paper specifically examines the impact of these factors within Chinese technical firms, focusing on technology enterprises like Qihoo, Didi global and Ctrip.com. Organizational performance in these companies is conceptualized as the integration of AI-driven knowledge sharing, knowledge-based organizational support (KOS), Knowledge Management Dynamic Capabilities (KMDC), and organizational learning, alongside structured training initiatives. These elements collectively aim to improve firm revenue, enhance stakeholder experiences, and ensure business continuity, as supported by previous research (Gigauri, 2020). Furthermore, the study adopts the concepts of knowledge management (KM) and knowledge-based organizational support, defined as processes involving the generation, storage, sharing, and application of knowledge resources (Di Vaio et al., 2021). By leveraging these practices, Chinese technological companies can align their operational frameworks with strategic goals, fostering long-term growth and resilience. The theoretical foundation of this research is grounded in the Resource-Based View (RBV) and Knowledge-Based View (KBV), which offer robust frameworks for analyzing the dynamic interplay of knowledge resources and capabilities within organizational contexts. These models provide critical insights into how organizations can optimize their performance by effectively deploying and integrating AI technologies, emphasizing the strategic importance of knowledge as a key organizational asset.

The Resource-Based Theory (RBT) posits that superior firm performance hinges on a firm's ability to possess and effectively deploy a unique set of strategic resources (Barney, 1991; Barney et al., 2011). Alongside this, the Technology Acceptance Model (TAM) (Davis, 1989) provides a complementary framework by emphasizing the role of perceived usefulness and ease of use in the adoption of innovative technologies. The theoretical underpinnings of these frameworks are integral to this study. Moreover, The Resource-Based View (RBV) serves as a foundational framework in strategic management and organizational studies, exploring how firms sustain competitive advantage by leveraging distinctive and valuable resources, such as knowledge management, knowledge sharing, and organizational learning. Originating from the work of Jay Barney in the 1980s, RBV has become a cornerstone in understanding how firms achieve long-term success. Similarly, the Knowledge-Based View (KBV), developed primarily by Fred Davis in the late 1980s, has been instrumental in the fields of information systems and technological adoption research. KBV extends RBV by specifically addressing the role of knowledge as a critical resource for innovation and performance. Our study integrates RBV and KBV to analyze how Chinese technological firms—such as Chinese Qihoo, Didi global and Ctrip.com—utilize knowledge sharing, artificial intelligence (AI) technologies, organizational learning, and knowledge management processes to enhance firm performance. Specifically, it explores the mechanisms by which knowledge-based organizational support (KOS) fosters competitive advantages and improves organizational outcomes (Gunjal, 2019). The processes of knowledge management (KM) are conceptualized into four constructs: knowledge generation, sharing, storage, and utilization (Abudaqa et al., 2021). Knowledge generation, for instance, is defined as the acquisition and development of knowledge tailored to meet current needs while acting as a resource for future applications (Antonelli & Colombelli, 2018). By synthesizing these theories, the study provides a comprehensive framework for understanding how KM and organizational learning collectively drive performance in Chinese technological enterprises.

Our paper makes three contributions to the literature. First, we contribute to the organizational performance literature. Chinese technology companies' knowledge sharing with AI, organizational learning, and KOS, KMDC is the inspection and management of the realization of organizational performance and goals. According to the above arguments, it can be a company, a team, or a function. Depending on the level of inspection, it can also reflect the overall business development and business performance of the organization. Indeed, Chinese technology companies' organizational performance mainly reflects the overall performance of the organization and mainly assesses the organization's profits and revenues. The Chinese technology company's key performance indicators, company-level key tasks, completion of organizational performance contracts, and overall business development of the organization. Organizations are transforming into knowledge-intensive facilities through adaptive, hierarchical, and generative learning direction (Herremans & Isaac, 2005). The Chinese technology companies' knowledge sharing important and vital information among and across the organization for the development and success of the company and industry. In other words, Variables like OL also utilize knowledge resources and management systems and inculcate them in a way that organizational performance is ensured (Camisón et al., 2017). Our research will examine the impact of knowledge management, knowledge sharing and organizational learning on organizational performance from Chinese technology companies. It includes share different knowledge, documents, and codes of the enterprise to continuously improve the technical capabilities of other meta-functions. Secondly, our study 's other important constituents and affecters of this relationship include Organizational learning (OL) and KMDC, KOS practices and orientation and acceptance of external information sources. Prior studies focus on knowledge factors and organizational Learning; Besides, Organizational learning (OL) is the formalized process of knowledge creation, retention and transference between individuals, teams, executives of an organization (Baporikar, 2020). From employee learning behavior undertaken by an organization, the tendency to commit errors and mistakes diminishes over the passage of time and impactful knowledge resources are created. Accordingly, these reserves encompass the varied domains of the organization and facilitate employees of every level in learning new competencies and attaining low levels of errors (Darwish et al., 2018). Lastly, our study expands the understanding of the knowledge management dynamic capacities in the digital era. Moreover, knowledge management dynamic capacities (KMDC) are also important for these companies. Generative knowledge management dynamic capacity refers to the ability to discover or improve knowledge and the processes, technologies, software products, and services that derive from it. Moreover, it is based on the system's intellectual and creative capital, which is present among its members, research infrastructure and alliances. Likewise, disseminative capacity denotes the ability to contextualize, format, adapt, translate, and diffuse knowledge through a social and/or technological network and to build commitment from stakeholders. meanwhile, the dynamic implications of the model suggest to the need for constant vigilance and adaptation in the midst of an economy characterized by rapid change. Knowledge Dynamic capabilities thus reflect an organization's ability to achieve new and innovative forms of Chinese technology companies's competitive advantage (Parent, R., Roy, M., & St-Jacques, D. (2007)).

The research questions guiding this study are as follows:

**RQ1:** To what extent do knowledge-based organizational support (KOS), knowledge sharing facilitated by AI technologies, continuous organizational learning, and knowledge management dynamic capability (KMDC) positively influence the organizational performance (OP) of Chinese technological firms, specifically Chinese technology companies?

**RQ2:** How do staff skills, education levels, and technological innovation moderate the relationships between KOS, knowledge sharing with AI technologies, continuous organizational learning, KMDC, and the organizational performance of Chinese technological companies?

These questions aim to explore the interplay between organizational knowledge frameworks and digital innovation while considering the moderating effects of human and technological factors. Indeed, this paper intends to extend the literature in this area by examining the knowledge-based review theory and RBV theory, and measure the organizational performance of Chinese technological company's multidimensional aspects of knowledge-based organizational support, knowledge Sharing, KMDC and organizational learning and their impact on organizational performance and product release in China. Besides, our findings of the study can provide insights for organizations in the Chinese software industry to improve their profits and organizational performance and project management through effective knowledge management and organizational learning initiatives. In this research study, we explore the impact of knowledge-based organizational support, knowledge sharing with AI technologies, OL, KMDC on organizational performance through Chat GPT, AI technologies and technology innovation of firms (Chatterjee, S., Rana, N. P., Tamilmani, K., & Sharma, A. (2021)). As a result, our study provides empirical evidence on how KOS, KMDC, Knowledge sharing with AI technologies, organizational learning and organizational performance influences organizational performance of Chinese companies, and we hope it will be useful for future research and practice.

*The remainder of this paper is organized as follows. Following a brief introduction in Section 1, a literature review and research model and hypothesis development are presented in Section 2. Section 3 introduces the research methodology, including the questionnaire design, data collection, and reliability and validity PLS-SEM testing. Section 4 adopts a structural equation modeling approach to test the hypotheses and emperical analyzes the results based on an empirical study. Finally, Section 5 summarizes the research findings and discusses future research directions.*

## 2. Literature Review

Our study also investigates KOS, knowledge sharing with AI technology, organizational learning, and KMDC as precursors of organizational performance of Chinese technology companies. Moreover, the staff skills, technology innovation, and education level play a moderating role between KOS and the OP of Chinese's technological companies. Our paper uses knowledge base review and the RBV Model and KBV theoretical model for discussion. Using this approach, this article shows that the educational level of the employees, their technical ability, and technological innovation also affect organizational performance, especially AI technology. The application of chat GPT technology can have a continuous impact on the development of the Chinese firm's organizational performance.

Our study investigates the impact of organizational performance and related knowledge factors using a research theoretical model and combining Chinese technology companies. This study mainly refers to the KBV theory and RBV theory model for analysis and research. The KBV and RBV theory is aimed at explaining and predicting the firm's acceptance and adoption of new information technologies by users. It is particularly effective in understanding user behaviors and attitudes towards the adoption of different technological innovations. In addition, another theoretical model, the RBV theory model, it was also used in this article for the analysis of research. The Resource-Based View (RBV) is a theoretical framework commonly used in strategic management and organizational studies. Its focus is on the understanding of how firms can gain and maintain competitive advantage through the use of the unique and valuable resources that they possess. Developed primarily by Jay Barney in the 1980s, RBV theory has become a fundamental concept in the field. After that, this paper also explores the changes in organizational performance in China, which is a Chinese cybersecurity technology company, through research and analysis using the above two theories in combination with the subject research in the paper. The impact and improvement of knowledge sharing, KMDC, organizational learning, and AI technology organizational performance.

Furthermore, Chinese technological companies' knowledge-based organizational support and collaboration management (KOCS) may be defined in several different ways based on the context in use. Beesley and Cooper, for instance, attempted to find a consensus on how knowledge and KM should be defined. Their investigation on the subject found that the most cited definition used to describe knowledge is given by Polanyi (1966) who distinguished two fundamentally different kinds of knowledge sharing with AI and knowledge management. According to Polanyi (1966), the first type of knowledge is called explicit knowledge management, which is personal/individual of Chinese technology companies. This type of knowledge can be expressed in forms such as written words, excel, PowerPoint, pdf file, software and code management, pictures, or numbers, and it can be communicated and transferred knowledge information quite easily. The second type of knowledge management is known as tacit knowledge and AI technologies, which is highly personal summaries and difficult to properly formalize, communicate, and easily share with others. Consequently, as for the term "Knowledge management (KM)," In the field of knowledge management, there were several proposed models. However, based on our extensive literature review, the most commonly-used model was the one formulated by Marquardt (Marquardt, M) and Beesley and Cooper (Beesley, L.; Cooper, C), whose KM model includes four main four components:

(1) Knowledge acquisition of Chinese technical firm.
(2) Knowledge creation of Chinese technical firm.
(3) Knowledge storage and retrieval of Chinese technical firm.
(4) Knowledge transfer and utilization with AI technologies of Chinese technical firm.

In this study, this knowledge-based organizational support variable theory model also adopted Marquardt's and Beesley and Cooper's KM models. Therefore, the first hypothesis is as follows:

*H1. Knowledge-based organizational support (KOS) has a significant positive effect on the organizational performance of Chinese technology companies (OP).*

Furthermore, Chinese technology companies' organizations also were transforming into knowledge-intensive facilities and organizational learning through adaptive, knowledge, AI technologies, and transfer, hierarchical, and generative learning direction (Herremans & Isaac, 2005). This study found that continuous learning by organizations and corporate employees can improve the overall performance of the company. The Organizations' knowledge of Chinese technology companies is important and vital information among and across the organizations accounts for the development and success of the company and industry. Thus, measure variables like organizations learning also utilizes knowledge resources and management systems and inculcates them in a way that organizational performance is ensured (Camisón et al., 2017). Therefore, we propose the following hypothesis:

*H2. Continuous learning and training of employees OL has a significant positive effect on organizational performance of Chinese technology companies (OP).*

Furthermore, Chinese technological companies' knowledge sharing with AI technologies is the key to opening knowledge management. AI technologies, and knowledge sharing with AI technologies is constantly associated with the strategy to compete in maintaining an organization's core competence and competitive advantage (Alavi & Leidner, 2001). It also includes AI, and ChatGPT technologies. Nevertheless, knowledge sharing and AI technologies under a certain circumstance is considered unreasonable since the knowledge owned by individuals is deemed as valuable assets. Indeed, *Chinese technology companies* have developed some knowledge-sharing platforms and software using AI technology, such as Chat GPT. *Chinese technology companies'* employees can generate AI knowledge, knowledge codes, and knowledge contexts according to different employee needs, and the generated content can be shared with different employees. Hence, these individuals of *Chinese technology companies* are more likely to keep the information they own to secure their positions in an organization. Knowledge sharing with AI technologies is important in knowledge management process, which gradually improves and fixes the production system and relevant elements (Darudiato & Setiawan, 2013). As a result, knowledge sharing is tightly related to a company's long-term performance and competitive advantage. Therefore, we propose the following hypothesis:

*H3. Knowledge Sharing (KS) with AI Technology has a direct positive effect on organizational performance of Chinese technology companies (OP).*

Additionally, the idea of dynamic capability originated in the strategy field and was encapsulated in the classic paper. This has spawned a number of papers, which have formed a dominant perspective, although in recent years the idea of dynamic capabilities has also been adopted by functional disciplines such as *Chinese technology companies'* marketing, human resources and information technology capabilities. Therefore, this study starts by reviewing some key debates within the 'dominant' tradition, and then summarize some of the perspectives and contributions from functional disciplines. We finish the section with a summary of the main areas of agreement and disagreement with regard to dynamic capabilities (Easterby-Smith, M., & Prieto, I. M. (2008)). Therefore, we propose the following hypothesis:

*H4. Knowledge management dynamic capabilities KMDC has a significant positive effect on organizational performance of Chinese technology companies (OP).*

What's more, the study measurement scale for Staff Skills, Technology Innovation, and Education level is composed of three items adapted from Dess and Beard (1984) and Mintzberg (1979). Technological innovation includes the use of the latest R&D technologies, opensource or self-programming frameworks, and GitHub open-source codes, as well as the use of the latest innovative technologies, such as generative AI, big data models, and other technologies. Employee skills include, for example, employee code quality, employee work efficiency, improvement of employee work information skills, etc. moreover, the successful release of each project by an Internet company also plays a very important role in improving the company's organizational performance. Meanwhile, whether the project is delayed or released in advance also indicates whether the product can be released successfully. The respondents were asked the degree of perceived unpredictability and complexity, such as "the external environmental change in our industry is severe," "our customers frequently request new products or services" and "in our software industry, there is a constant demand for adaptation to firm change." Therefore, we propose the following hypothesis:

*H5. Education Level (EL), Technology Innovation (TI), Staff Skills (SS) with technology has a direct positive effect on organizational performance of Chinese technology companies (OP).*

Consequently, Chinese technology companies' organizational performance is the ability of an organization to reach its goals and optimize results. The organizational performance results include firm's profits, revenue, and Enterprise work efficiency, product delivery speed, etc. In today's workforce, '*Chinese technology companies'* organizational performance can be defined as a company's ability to achieve goals in a state of constant change. Moreover, one of the important research questions in business has been why some organizations succeeded while others failed. Organization performance has been the most important issue for every organization be it profit or non-profit one. Moreover, it has been very important for managers to know which factors influence an organization's performance for them to take appropriate steps to initiate them. However, defining, conceptualizing, and measuring performance have not been an easy task. Researchers among themselves have different opinions and definitions of performance, which remains to be a contentious issue among organizational researchers. The central issue concerns with the appropriateness of various approaches to the concept utilization and measurement of organizational performance (Venkatraman & Ramanuiam, 1986). Therefore, we propose the following hypotheses:

*H6. Education Level (EL), Technology Innovation with Code framework (TI), Staff Skills (SS) moderates the relationship between KOS, KS, KMDC, OL and OP of Chinese technology companies (OP).*

In short, researchers among themselves have different opinions of organizational performance, In fact, Chinese technology companies continues to be a contentious issue among knowledge and organizational researchers. For example, according to Javier (2002), performance is equivalent to the famous 3Es theory (economy, efficiency, and effectiveness) of a certain program or activity. However, according to Daft (2000), the organizational performance is the organization's ability to attain its goals by using resources in an efficient and effective manner. Quite like Daft (2000), Richardo (2001) defined organizational performance as the ability of the organization to achieve its goals and objectives. And improve firm's profits and delivery products successful. Then, Chinese technology companies' Organizational performance has suffered from not only a definition problem, but also from a conceptual problem. For education level, technologies innovation and employee skills address all decisions related to the selection and assignment of features to a sequence of consecutive Chinese technology companies' product releases. In particular, the application of Pfleeger's (Regnell, B., Beremark, P., and Eklundh, O) definition of good decisions in software engineering, this study assume that a good release plan should be characterized by the following criteria:

It provides maximum Chinese technology companies' business value by offering the best possible blend of features in the right sequence of software product releases.

It satisfies the most important of Chinese technology companies' stakeholders.

It is feasible in terms of the existing Chinese technology companies' resource capacities available and existing dependencies between Chinese technology companies' features.

The Research Model and Hypothesis development - Conceptual framework: previous studies have based their criteria for selection on some knowledge factors and organizational performance, However, it had no AI technologies and control variables to impact organizational performance. Moreover, this research mainly focuses on knowledge based organizational support (KOS), KMDC, organizational learning, Knowledge sharing with AI technologies and organizational performance in Chinese small enterprises. On the other hand, this study also describes the impact of employees' educational background, technical capabilities, and corporate technological innovation on organizational performance. For example, highly educated employees tend to be more capable and are more efficient in product delivery and work efficiency for the company. Employees with strong skills are more effective in improving organizational performance. The company's product delivery and cost savings are also very important. The company's technological innovation, such as the number of patented technologies, researched internal performance platform software, efficiency tools, and other R&D innovations, will also improve the Chinese company's overall organizational job performance and knowledge management delivery, ultimately improving the overall efficiency and profits of the technological enterprise. Therefore, this study 's original data collection come from primary survey sources.

## 3. Methodology

### 3.1. Research Design

Our study focuses on KBV theory and RBV theory model. Likewise, this study defines a new research model in this article as a knowledge factors, AI technologies and organization performance improvement that was established within the past series years and offers an innovative product, education level of Chinese technology companies' employees, staff skills improvement, product service, or business model (Candi & Saemundsson, 2008; Zahra, 11295).  our study collected our data through an open-source online survey tool (Microsoft online survey) from My former colleagues' WeChat group using a simple random sampling technique and eclectic questionnaires. Indeed, based on the relationship between the factors of knowledge support, organizational learning, knowledge sharing with AI technologies and KMDC and organizational performance and Chinese technology companies' profits and product delivery, and above analysis and introduction, we can propose the following theoretical framework model. Knowledge sharing with AI is being used in different sectors and discuss the Organizational performance challenges and opportunities they pose. Thus, the theoretical conceptual framework, Figure 1 is presented below.

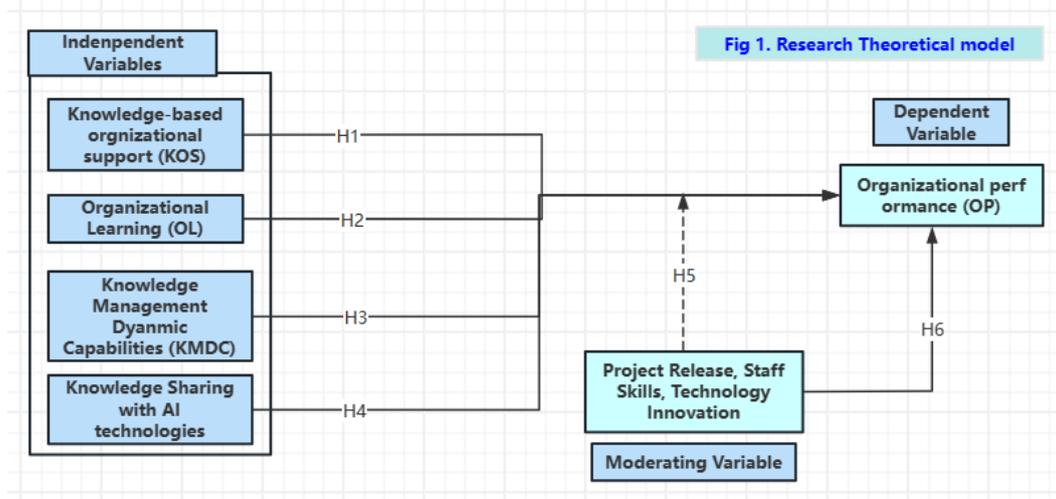

**Figure 1.** The Conceptual framework (Source: Authors' own work)

*3.2. Data and Sample Selection*

The data for this study were primarily derived from a survey conducted by the author, with responses collected from 129 employees across firms (Chinese Qihoo, Didi global and Ctrip.com). The survey employed electronic questionnaires distributed via WeChat and WhatsApp groups, primarily targeting colleagues from public sector organizations who had adopted digital technologies to enhance their operations and service delivery. A total of 129 valid responses were obtained and subsequently analyzed. Moreover, to test the proposed theoretical research model, a Partial Least Squares (PLS) analysis was conducted. The survey items were measured using a five-point Likert scale, ranging from (1) "Strongly Disagree" or "Neutral" to (5) "Strongly Agree" or "Always." The questionnaire was designed using Microsoft's online survey platform and distributed to former colleagues from Chinese companies via WeChat groups to ensure comprehensive and relevant data collection. As a results, our final dataset comprised 129 observations from employees of Chinese companies, representing a valuable sample for examining the study's objectives. To analyze the data, this study employed SPSS and SmartPLS 4.0 software, utilizing quantitative methods such as linear regression and empirical analysis. These analytical techniques were applied to verify the research hypotheses, thereby enabling the study to draw robust and evidence-based conclusions.

*3.3. Measures and constructs*

Measurement items for this study were developed based on established scales from the existing literature, with adjustments made to adopt a team-based referent-shift format where applicable. As outlined in Appendix A, the study measures key variables—including organizational performance, knowledge sharing with AI technologies, organizational learning, and control variables such as technology innovation, employee skills, and education level—at the organizational level, focusing on Qihoo, Didi global and Ctrip.com in China.

The six variables were assessed using a five-point Likert scale, ranging from "Strongly Agree" to "Strongly Disagree," with multiple measurement items for each construct. The measurement items and their sources were carefully selected and adapted from prior studies to align with the study's marketing and technological contexts. Adjustments were made to ensure the measures were suitable for the online survey format used in data collection.

Furthermore, Appendix A provides a comprehensive summary of the 34 metrics employed in the study. Each questionnaire item was designed to capture the nuanced aspects of the constructs under investigation, ensuring validity and reliability. These measures were refined to reflect the specific context of Chinese technological firms, thus enhancing their applicability to the study objectives.

## 4. Results and Data Analysis

### *4.1. Characterise Demographic*

According to above description, we explain this study's approach to the research results and describe exactly what steps you will take to Graph Chart and Data table as following (as shown in Figure 2).

Figure 2. Proportion relationship chart of employees with different Gender and Education level in Chinese companies.　(Source: Authors' research work.)

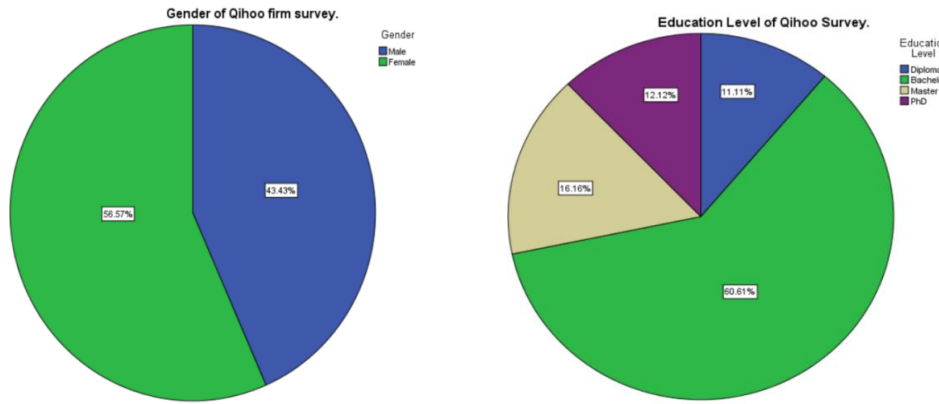

According to the figure above, we conducted an analysis in which 43.43% of respondents are male and 56.57% are female. Women make up most respondents in this survey. In addition, different educational backgrounds also have an impact on organizational performance. At 60.61%, we can see that the highest proportion of respondents in the survey data had a bachelor's degree. Employees with other academic qualifications, such as doctorates, masters, and diplomas, accounted for 12.12%, 16.16%, and 11.11%, respectively. We can see that in Chinese technology companies' survey on organizational performance change, the proportion of employees with a bachelor's degree accounted for most employees. In additional, Chinese technology companies' employees with different roles in the enterprise are also very important to the organizational performance and knowledge management of the Chinese technology companies' organization, and employees with different roles have different levels of attention.

Figure 3. Proportion relationship chart of employees with different roles in Chinese technology companies agreeing on the role of KOS, KS, OL, KMDC, SS, TI, Education level on firm's organizational performance.

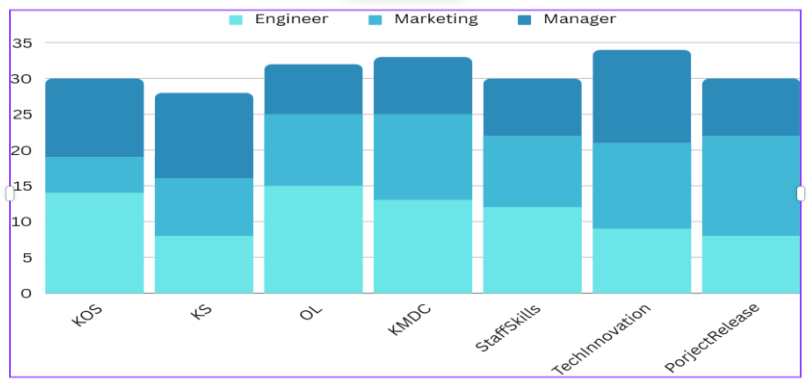

Notes. Different role of job performance (Source: Authors' research work)

According to Figure 3 analysis, among them, The Chinese technology companies' employees in the roles of engineers, Marketing, Operation, and team managers have a higher recognition of the use of technological innovation, and KMDC, KOS, Knowledge sharing with AI technologies and Organizational Learning have a higher recognition of the improvement of the Chinese technology companies' organizational performance. That is, the understanding of the dynamic capabilities of technological innovation, organizational learning, and KMDC knowledge management were more important factors. The following are data descriptive statistics, and the statistical results of the relationship analysis between various variable items are as follows:

**Table 1.** Descriptive Statistics.

| Descriptive Statistics | | | | | | |
|---|---|---|---|---|---|---|
| Variables | N | Minimum | Maximum | Mean | Std. Deviation | Variance |
| KOS | 129 | 1.0 | 5.0 | 4.152 | .8374 | .701 |
| KS | 129 | 1.0 | 5.0 | 4.061 | .8308 | .690 |
| OL | 129 | 1.0 | 5.0 | 4.131 | .8036 | .646 |
| KMDC | 129 | 1.0 | 5.0 | 4.141 | .8923 | .796 |
| OP | 129 | 1.0 | 5.0 | 4.182 | .9727 | .946 |
| TI | 129 | 1.0 | 5.0 | 4.222 | .7222 | .522 |
| EL | 129 | 1.0 | 5.0 | 3.8129 | .8863 | .786 |
| SS | 129 | 1.0 | 5.0 | 4.141 | .8453 | .714 |
| Valid N (listwise) | 129 | | | | | |

Note: Standard errors are in parentheses; p < 0.1, ∗ p < 0.05, ∗∗ p < 0.01, ∗∗∗ p < 0.001, respectively. (Source: author's work)

From above Table 2, and Figure 2 and Figure 3, As we can be seen from the above Bar chart, the data collected by our research questionnaire mainly comes from the well-known Chinese technology companies, and the respondents to the questionnaire also come from different Chinese technology companies' departments survey observations.

Table 2. The Correlations Matrix.

| Variable | KOS | KS | OL | KMDC | OP |
|---|---|---|---|---|---|
| KOS | 1 | | | | |
| KS | .823** | 1 | | | |
| OL | .668** | .829** | 1 | | |
| KMDC | .859** | .814** | .685** | 1 | |
| OP | .742** | .719** | .596** | .6129** | 1 |

Note: **. Correlation is significant at the 0.01 level (2-tailed). (Source: Author's work)

Note: Standard errors are in parentheses; p < 0.1, ∗ p < 0.05, ∗∗ p < 0.01, ∗∗∗ p < 0.001. respectively. (Source: author's work)

Figure 4. Correlation Coefficient Matrix and Regressions Model Trend Graphic.

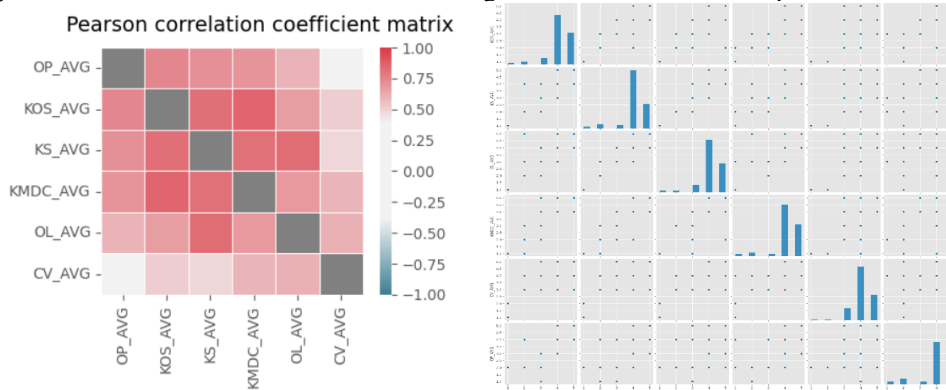

(Source: author's work)

From above table 2 and Figure x's Chart and data description, we conducted visual and statistical analyses of the regression design. the statistical results based on above data correlation analysis are as follows. this study sorted out the analysis data collected by the questionnaire, calculated the average value e as the data for Independent Variable, dependent Variable data analysis, and conducted multivariate linear regression analysis for different variables, and obtained the following data analysis results; furthermore, for above statement analysis, we can know that because of the independent variable 's P-value < 0.05, and all IV variables' correlations value >0.00. So, we Rejected the null and accept the alternative. Moreover, the statistics and correlation were statistically significant. therefore, there was a linear relationship between Organizational learning, Knowledge sharing, Knowledge Management Dynamic capabilities, Knowledge-based organizational support, and Organizational performance of Chinese technology companies. Correspondingly, the Knowledge sharing with AI Technologies,

Knowledge Management Dynamic capabilities, Knowledge-based organizational support, Organizational continue Learning and knowledge training, and the all-independent coefficient of determination (r2) is. .742**, .719**, .596**, .6129**.   Overall, there is a positive relationship between organizational learning, knowledge sharing with AI technologies, knowledge management dynamic skills, knowledge-based organizational support, and organizational performance.

*4.2. Validity and Reality Analysis*

According to this research study model and hypothesis, this study also came up with the following research equation. in this model, this equation shows the relationship between the research model and the regression analysis of the dependent, control variables and independent variables. Because we have multiple factors for most of the organizational performance of Chinese technology companies in our study, we use a fixed effects specification. Therefore, we estimated the following Model 1 's equation (1):

$$Y_{op} = \alpha + \rho_1 x_{kos} + \rho_2 x_{kmdc} + \rho_3 x_{ol} + \rho_4 x_{ks} + Control(x_{education}) + \partial \quad (1)$$

Based on the hypothesis development, this study model can be illustrated as shown.
We estimate the following Model 2 's equation (2):

$$Y_{op} = \alpha + \rho_1 x_{kos} + \rho_2 x_{kmdc} + \rho_3 x_{ol} + \rho_4 x_{ks} + Control(x_{technology}) + \partial \quad (2)$$

We estimate the following Model 3 's equation (3):

$$Y_{op} = \alpha + \rho_1 x_{kos} + \rho_2 x_{kmdc} + \rho_3 x_{ol} + \rho_4 x_{ks} + Control(x_{staffskills}) + \partial \quad (3)$$

Where $Y_{op}$ is organizational performance of Chinese technological firm, and then the coefficients β1, β2, β3 and β4 represent the efficiency of every variable factor on organizational performance of Chinese technology companies.  $x_{kos}$ is the variable of Knowledge-based organizational support of firm, $x_{kmdc}$ is Knowledge management dynamic capability of firm, $x_{ks}$ is  knowledge sharing of firm, $x_{ol}$ is organizational learning of firm, and $x_{education}$, $x_{technology}$, $x_{staffskills}$ is education level, technology innovation, and staff skills, $\partial$ is error constant. According to the above description, we see that the control variables include several variables, and this study can derive the following equation (X).

$$CV_{control} = X_{education} + X_{technology} + X_{skills} \quad (X)$$

Where $CV_{control}$ is control variable, $X_{education}$ is Education level of employee, $X_{technology}$ is technology innovation of employee, $X_{skills}$ is firm employee's skills.

Furthermore, as the organizational performance of Chinese technology companies improved or increased, organizational learning, knowledge sharing, product release, knowledge management dynamic skills, knowledge-based organizational support KOS, and firm organizational learning also improved or increased. To sum up, this study knows that the there is a linear relationship between Organizational learning, Knowledge sharing with AI, Knowledge Management Dynamic capabilities, Knowledge-based organizational support (KOS), Organizational Learning and the Education level, Staff skills, Technology innovation, and Organizational performance of firm. Thus, according to the above data description and analysis, the research statistics and correlation are statistically significant. there is a linear relationship between knowledge-based organizational support and collaboration and to Chinese technology companies. the coefficient of determination (r2) is greater less positively number. So, Chinese technology companies' digital knowledge-based organizational support and team collaboration also has an impact on Chinese technology companies' organizational performance. Thus, all hypothesized suggestion is supported. The following are the empirical analysis results based on the above three different three model equations.

**Table 3.** The Organizational performance and Knowledge factors statistic and empirical analysis.

| Variables | Model 1 (OP) | Model 2 (OP) | Model 3 (OP) |
| --- | --- | --- | --- |
| KS With AI | .841* | .811 | .835 |
|  | (10.177) | (8.134) | (11.904) |
| KOS | .731 | .765 | .721 |
|  | (10.917) | (8.904) | (10.317) |
| KMDC | .731 | .834 | .731 |
|  | (9.221) | (11.221) | (9.451) |
| OL | .721 | .834 | .702 |
|  | (.721) | (12.221) | (8.958) |
| Education Level | .765 |  |  |
|  | (10.171) |  |  |
| Technology innovation | | .635 (11.221) | |
| Staff skills | | | .725 (10.131) |
| Constants | .765 (10.171) | .1295 (11.131) | .801 (9.201) |
| Observation | (N=129) | . (N=129) | (N=129) |

Note. t statistic is in parentheses; p < 0.1, ∗ p < 0.05, ∗∗ p < 0.01, ∗∗∗ p < 0.001., respectively. (Source: Authors' own Work)

As can be seen from Table 3 above, the coefficient of determination (P value and T value) is a measure of the predictive power of a research model, indicating the research proportion of variance in an endogenous construct explained by its predictor different variables. And the coefficient and T-value of knowledge sharing (the IV variable) with artificial intelligence technology are both positive values. This indicates that AI technology has a positive effect on knowledge analysis, knowledge management, and organizational learning. There is a positive correlation with the Chinese technology company, (DV variable). From the above three models, we can see that employees' educational background, employees' technical ability, and the firm's technological innovation ability are also positively related to the final organizational performance because the P value = 0.000 < 0.05 and the coefficient values are all positive. There is a positive relationship; the coefficients are all positive, and the P value is 0.001<0.05. Therefore, we will reject zero and accept the alternative. Therefore, the proposition of this hypothesis is supported. Control variables (IV variables) They have a positive relationship with the organizational performance of Chinese technology companies (DV variables). From the above three models, we can see that the coefficients of employees' educational background and technical ability are larger. In other words, employees with a higher educational background have a better effect on improving the final performance of the company. The coefficients of technical ability and educational background are also very large, meaning that technical ability and educational background are equally important and play a great role in improving Chinese technology companies' organizational performance and organizational effectiveness. So, this hypothesized suggestion is supported.

The following are the results of the reliability analysis and Validity analysis Results.
Table 4. Reality and Validity Items (N = 129).

| Construct items | Cronbach's alpha | CR | AVE |
|---|---|---|---|
| Organizational Learning | 0.895 | 0.900 | 0.750 |
| Knowledge Sharing & AI | 0.898 | 0.901 | 0.752 |
| Knowledge-based organizational support | 0.882 | 0.896 | 0.743 |
| Knowledge management Dynamic Capability | 0.832 | 0.882 | 0.714 |
| Control Variables | 0.873 | 0.894 | 0.737 |
| Organizational Performance | 0.924 | 0.908 | 0.768 |

Note: The Validity and Reality results of KOS, KS, OL, KMDC, Control variables and OP, respectively. (Source: Author's own work)

According to the data in the above table 4, we can know that all variable values of Cronbach's alpha, CR, and AVE are consistent with the hypothesis test, supporting the establishment of all the above hypothesis tests. This research study detects questionnaire validity through internal consistent and convergent validity criteria. According to internal consistency, Cronbach's alpha should yield values over 0.6, and valid items should yield values over 0.7 (Nunally, 1967). All items were measured using a five-point Likert scale (1 = strongly disagree and 5 = strongly agree). Table X shows all items present a value greater than 0.8. There are four measurement models in this study: (1) Knowledge-based organizational support (KOS), (2) knowledge management (KM), (3) organizational continue learning and training (OL), (4) Knowledge Management Dynamic Capability (5) Control variables and (6) sustainable organizational performance (OP). Verification of each measurement model was as follows. Firstly, we performed PLS-SEM equation analysis using the Smart-PLS on some indicators measuring the level of practice of Knowledge-based organizational support, knowledge management, Control variables, Knowledge Management Dynamic Capability, organizational learning, and sustainable organizational performance. Secondly, according to Hair et al. [100], the reliability of the measurement models should be first verified using Cronbach's alpha or composite reliability (CR) to evaluate the construct measures' internal consistency reliability, followed by the validity assessment of the measurement models.

### 4.3. The structural measurement results.

After the reliability and validity evaluation of the measurement models was confirmed in Chinese technology companies' observations. Moreover, assumed relationships presented in the structural model can be assessed using PLS-SEM technique. The structural model assessment started with the determination of path coefficients($\beta$), coefficients of determination ($R^2$), and effect size ($f^2$). The results of the hypotheses tests using PLS-SEM model equation are presented in Table 5 and Figure 5.

**Figure 5.** Structural Model Evaluation and Hypotheses test Results.

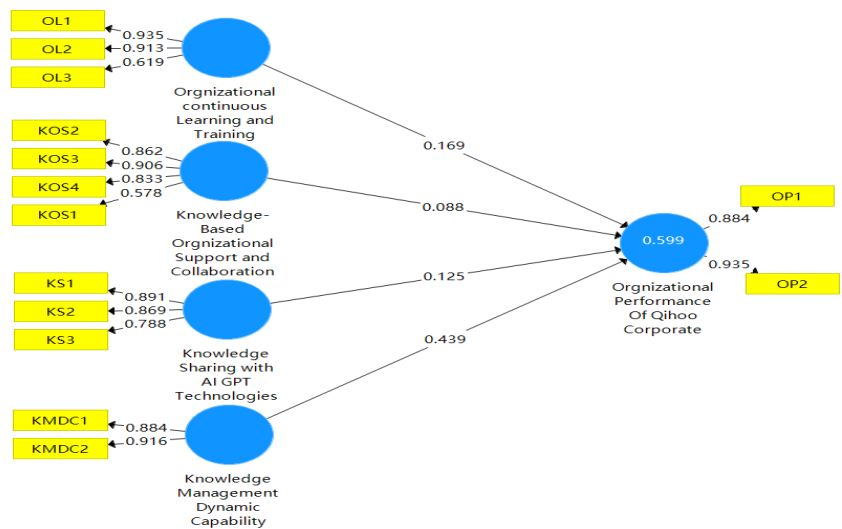

(Source: Authors' research work)

Furthermore, according to above figure5 and table 5 relationship results of different hypothesis description. The PLS-SEM path coefficients, also known as standardized regression coefficients, provided estimates of the relationships between constructs in the structural model. These coefficients indicate the strength of strength, direction, and statistical significance. The stages in the outer model design using PLS algorithm analysis in Figure 5 and Table 4 used convergent validity to determine the validity of each relationship between indicators and their latent variables by examining outer loadings. Loading factor values were greater than 0.7 and considered valid when the average variance extracted (AVE) value was above 0.5 (Chin, 2009). Organizational Continue learning, Knowledge factors and AI technologies, as well as organizational performance variables, had average outer loadings above 0.7, showing that all indicators fulfilled convergent validity with high values. The AVE values were above 0.50, indicating that all variables had good convergent validity (Sekaran & Bougie, 2016). Moreover, Composite reliability for each variable was above 0.7, hence the instrument possessed high consistency and precision in measurement (Hair et al. 2014). The Cronbach's alpha values average above 0.6 for all variables showed the reliability of the instrument (Ghozali, 2016). Based on the convergent validity, composite reliability, and Cronbach's alpha, the indicators within each variable could be used for further analysis (Hair Jr et al., 2014; Leguina, 2015). Meanwhile, presented as standardized values ranging from -1 to +1, higher absolute values represent stronger (predictive) relationships between constructs, while values closer to zero indicate weaker relationships. Moreover, the significance of the path coefficients is assessed using the bootstrapping method. Based on the above empirical analysis, term relationship analysis, regression analysis, and PLS -SEM equation analysis, we can draw the following conclusion: Among this analysis, the impact of AI technology, as well as some control variables of model — employees' educational background, technical capabilities, and the Chinese technological innovation— are the contributions of this finding. Finally, our study also can be concluded that this hypothesis has been supported as per the results of the study.

**Table 5.** The Results of Direct Relationship.

| Hypotheses | Relationships | Status | t-value | p-value | Decision |
|---|---|---|---|---|---|
| H1 | KOS->OP | Positive | 10.917 | .000 | supported |
| H2 | KS with AI -> OP | Positive | 10.177 | .000 | supported |
| H3 | OL->OP | Positive | 7.304 | .000 | supported |
| H4 | KMDC->OP | Positive | 7.304 | .000 | Supported |
| H5 | Control Variables->OP | Positive | 3.152 | .000 | Supported |

Note. N = 129, *** for significance at p < 0.001, and ** for significance at p < 0.05 level, respectively. (Source: Author's own work).

After all, according to the results shown in Table 5, this study also had proposed that knowledge-based organizational support significantly positively influences organizational performance of Chinese technology companies (i.e., Hypothesis 1). Furthermore, the research results show that knowledge sharing positively affects organizational performance, supporting Hypothesis 2. Similarly, Hypothesis 3, which proposed that Organizational learning significantly influences organizational performance, was also supported. Similarly, Hypothesis 4, which proposed that knowledge management dynamic capability significantly influences organizational performance, was also supported. As expected, organizations' dynamic capability, organizational learning and knowledge sharing can help shape the change organizations performance to implement, as the literature suggests (Li et al., 2021). Besides, Hypothesis 5 and Hypothesis 6. proposed that staff skills, education level, and technology innovation also has a positive relationship with organizational performance. All in all, From the analysis results of the questionnaire data collected by Chinese technology companies, Thus, we can see that these digital factors have a positive relationship with digital transformation and All hypothesis tests were supported, respectively.

*4.4. Robustness analysis*

To summarize the key terminology, in order to improve the robustness of the test research model, this study used different sample data to conduct different model tests. according to table 7 empirical analysis and descriptive, and the results show that KS with AI, KOS, KMDC, and OL are positively related to organizational performance in some research models. And the firm employee's educational background and technological innovation are also significantly related. However, employee skills may sometimes not be as significant, i.e., firm's employee skills may not be one of the factors that have an impact on organizational performance. When the employee skills of a Chinese technical firm are included, KMDC may sometimes be insignificant, so employee skills may also sometimes have an impact on whether KMDC has a significant final impact on Chinese technology companies' performance. For robustness testing, the following equation supports;

Thus, we estimate the following Model 4 's equation (4):

$$Y_{op} = \alpha + \rho_1 x_{kos} + \rho_2 x_{kmdc} + \rho_3 x_{ol} + \rho_4 x_{ks} + ControlVariable + \partial \quad (4)$$

And about other robustness model, we estimate the following Model 5 's equation (5):

$$Y_{op} = \alpha + \rho_1 x_{kos} + \rho_2 x_{kmdc} + \rho_3 x_{ol} + \rho_4 x_{ks} + \partial \quad (5)$$

and other robustness model, we estimate the following Model 6 's equation (6):

$$Y_{op} = \alpha + \rho_1 x_{kos} + \rho_2 x_{kmdc} + \rho_3 x_{ol} + \rho_4 x_{ks} + ControlVariable + \partial \quad (6)$$

and other robustness model, we estimate the following Model 7 's equation (7):

$$Y_{op} = \alpha + \rho_1 x_{kos} + \rho_2 x_{kmdc} + \rho_3 x_{ks} + \partial \quad (7)$$

**Table 7.** The regression model of robustness analysis results.

| Variables | Model 4 (OP) | Model 5 (OP) | Model 6 (OP) | Model 7 (OP) |
|---|---|---|---|---|
| KOS | 0.392** | 0.469** | 0.344* | -0.089 |
|  | (2.65) | (2.84) | (2.62) | (-0.75) |
| KS | 0.228 | 0.332 | 0.380** | 0.809*** |
|  | (1.22) | (1.78) | (2.97) | (7.72) |
| OL | 0.214 | 0.018 |  |  |
|  | (1.40) | (0.13) |  |  |
| KMDC | -0.046 | 0.121 |  | 0.075 |
|  | (-0.31) | (0.80) |  | (0.69) |
| Tech Inno | 0.311* |  | 0.334** |  |
|         | (2.31)  |         | (2.75)  |         |
| Edu     | 0.323***|         | 0.311***|         |
|         | (3.89)  |         | (3.77)  |         |
| Skills  | -0.467***|        | -0.414***|        |
|         | (-4.59) |         | (-4.38) |         |
| Constant| 0.296   | 0.310   | 0.302   | 0.902***|
|         | (0.78)  | (0.86)  | (0.80)  | (3.73)  |
| Observations | 129 | 129     | 129     | 129     |

Notes. t statistics in parentheses. * $p < 0.05$, ** $p < 0.01$, *** $p < 0.001$, respectively. (Author's own work)

## 5. Discussion and Conclusions

In conclusion, by applying a KBV and RBV framework, our research presents a multi-factor conceptual model between KOS, KS with AI, OL and KMDC based on the perspective of organizational performance management, in which OP is measured using Five explanatory variables: KS, KOS, OL, KMDC and control variables. This study has defined the main knowledge and firm organizational performance concepts, this study investigates to understand the social and organizational factors that influence knowledge sharing with AI technologies. Consequently, the findings provide evidence for the importance of social capital as a lubricant of knowledge sharing and engaging performance management systems in knowledge-intensive organizations (Liao et. al., 2009). According to previous studies, we can see that there is no research on the impact of AI technology, corporate employees' academic backgrounds, technical capabilities, and corporate innovation on the ultimate performance of the company. This is the research gap in this study. Through this research, it was found that Chinese technology company – Chinese technology companies also have two contributions. The impact of AI technology, as well as the control variables—employees' educational background, technical capabilities, and the company's technological innovation—are the contributions of this finding. In short, this paper proposes some hypotheses to verify relationships among KMDC, Knowledge Sharing with AI technologies and AI tools organizational performance, and organizational learning (OL). Besides, Base on a sample of knowledge-intensive firms engaged in manufacturing, and financial sectors, data are collected using a mail survey, and hypotheses are tested using structural equation modelling (Liao et. al., 2009). Likewise, this study present Organizational Learning as a coordinating mechanism, and the results support it in these samples. Especially, knowledge management (KM) has recently emerged as a discrete area of organizations and frequently cited as an antecedent of organizational performance (Delen et. al., 2013). The analyses indicated that there is a strong and positive relationship between the implementation level of KM practices and organizational performance related to KM.

Our study aims to investigate the existing relationship between knowledge management (KM) infrastructures, knowledge management process capabilities, creative organizational learning (OL) and organizational performance (Bagheri et. al., 2015). This study's technology innovation, staff skills, and project release focus on the control role of KOS, KS, OL, and KMDC process for creative organizational learning and performance case study. Alternatively, prior studies that have noted the important of Chinese technology companies' organizational learning includes processes of training context creating, retaining, and transferring knowledge and has implications for the organizational performance and competitiveness of organizations. Moreover, Given the knowledge-based view of resources inherent in management of technology (MOT) (Argote et. al., 2017), in this study paper, it adopts an organizational learning framework that considers knowledge to be embedded in three major components of organizations – firm's members, tasks, and tools – and the networks formed by crossing them(Argote et. al., 2017). Alternatively, As mentioned in the literature review, Chinese technology companies' success requires high-performance human resource management practices (HRMP) and effective knowledge management capabilities (KMDC) to improve overall organizational performance of Chinese technology companies. Thus, only in this way can summarize corporate performance and product delivery be better improved (Gope et. al., 2018). Besides, our study examines the influence of OP on KMDC through a cross-case analysis including companies operating Chinese technology companies' different IT sector. In addition, Chinese firm's Organizational learning also aims to generate organizational learning in support of future strategic knowledge initiatives that will, in turn, foster knowledge asymmetries that can lead to differences in organizational performance.

**Practical Implications**

   Our results also have important practical implications for policy makers interested in organizations' performance and knowledge factors, and AI technology, based on the review of the literature and previous studies, the research model was generated to conceptualize the theoretical concepts and discover the research gap as well as to extend the theories of KBV and RBV through empirical evidence. First, this main goal of this study has been to provide insights into the characteristics of a knowledge factors and organizational Learning within organizations performance of Chinese companies. From a case study of a unique organization whose purpose is to facilitate strategic knowledge distillation, it was found that the process is characterized by targeted useful information gathering that relies on diverse experts for interpretation as well as validation. thus, this study also embodies the Chinese companies' organizational capability to leverage information technologies integrating them with processes for generating, storing, and transporting rich, deem bedded knowledge across multiple levels of the organization (Thomas et. al., 2001). In other words, Within the context of knowledge management, little research has been conducted that identifies the antecedents of a knowledge-cantered culture—those organizational qualities that encourage knowledge creation and dissemination. Second, The existing literature on organizational climate, job characteristics, and organizational learning (in the form of cooperative learning theory) is linked with the current thinking and research findings related to knowledge management to develop a theoretical model explaining the relationships among organizational climate, the level of cooperative learning that takes place between knowledge workers, and the resulting level of knowledge created and disseminated as measured by team performance and individual satisfaction levels (Janz et. al., 2003). Lastly, in comparison with previous research, our current study found that KOS, knowledge sharing with AI technologies, OL, and KMDC of Chinese technology companies is an antecedent of organizational management and organizational performance. project management and project performance play a mediating role between KMDC, OL, KOS, KS with AI technologies and the organizational performance of Chinese technology companies. In addition to the above, we can also see that the educational level of the Chinese technology companies' employees, the technical ability, and the technological innovation also have an impact on the organizational performance, especially the application of AI technology and GPT technology. the study basing on the relational view, resource-based view and the extended resource-based view highlights the benefits that accrue to Chinese technical firms that engage in knowledge sharing with AI technology, organizational learning, KMDC, and KOS to boost organizational performance of Chinese technology companies. All in all, these knowledge factors continuously affect the development of the organization's performance.

**Theoretical Implications**

   Our study makes three theoretical contributions. First, empirical study was conducted to verify the hypotheses by analyzing data from 129 valid questionnaires collected from manufacturing firms in China. based on the findings, Chinese Qihoo, Didi global and Ctrip.com companies want to narrow these organizational performance gaps and improve their organizational performance and productivity, they should create a work process or system that essentially helps strengthen the link between, KOS, KS with AI technologies, KMDC and organizational learning, so that they better adapt to the ever-changing market conditions and increasing competition. Meanwhile, this study provides opportunities for future research to explore more deeply the factors responsible for a lack of adoption and utilization of KOS, KS with AI technologies, KMDC, and OL in the construction knowledge management and organizational performance industry (Han, S.H., Yoon, D.Y., Suh, B., Li, B. and Chae, C. (2019)). Moreover, Chinese technology companies also was construction firms may better embrace policies that encourage employees to challenge the underlying assumptions behind the chronic problems of their day-to-day knowledge management, Knowledge Sharing, business operations and organizational performance improvement.
   Secondly, our study aims to find the informative value of focused on prediction organizational performance, knowledge innovation and KBV theory. The study contributes to a comprehensive understanding of the dynamics of knowledge management by examining the interplay between the dynamic capability of knowledge management (KMDC), knowledge sharing with AI technologies, knowledge-based organizational support, and organizational learning. This holistic approach provides a nuanced view of how these elements collectively influence organizational performance. On the other hand, this study highlights the role of AI technologies in promoting knowledge sharing within an engineering of Chinese technology companies. In particular, our finding contributes to growing literature on integrating AI into knowledge management processes, with practical implications for firms seeking to use AI technologies to improve knowledge exchange of firm. It links Chinese technology companies' learning and firm performance, contributing to literature on how learning initiatives affect overall firm performance. Our study also establishes a link between corporate learning and corporate performance, contributing to the literature on the impact of learning initiatives on business outcomes (Obeso, M., Hernández-Linares, R., López-Fernández, M. C., & Serrano-Bedia, A. M. (2020)).

Last, our study findings from the case study may be limited with respect to generalizing to other organizations. The specific context, industry, and organizational structure of Chinese technology companies may be very different from those of other technology companies. Consequently, by addressing these research gaps, this study also can contribute to the understanding of knowledge management, AI technologies, organizational learning, and organizational performance in the specific context of Chinese technology companies and other Chinese engineering companies. There may be a research gap in the in-depth exploration of the influence of external environmental factors, such as regulatory changes or industry trends, on knowledge management dynamics and organizational performance. It may lack an in-depth analysis of the longer-term effects of knowledge management initiatives on organizational performance (Han, S.H., Yoon, D.Y., Suh, B., Li, B. and Chae, C. (2019)). Understanding sustainable effects over a longer period of time could be a potential area for further research. Hence, our study is best way to have more than 2–3 companies as research objects. Due to the current limited time, this study only studies one company, which is Qihoo, Didi global and Ctrip in China. Chinese AI technology is not in-depth enough for internal knowledge management, knowledge sharing, and organizational learning and training in enterprises; more in-depth research and interviews are needed. Through the above data analysis and description, Chinese technology companies' organizational performance management continuously improves the overall organizational performance of Chinese technology companies through different knowledge management dynamic capability, knowledge sharing, and organizational learning activities. furthermore, the realization of organizational performance of Chinese technology companies should be based on the realization of individual performance, but individual achievement of performance does not necessarily guarantee that the organization is performing (Paul Hughes; Ian R. Hodgkinson). Furthermore, If the organization performance of Chinese technology companies is decomposed to each job and each person according to certain logical relationships, as long as each person meets the requirements of the organization, and the performance of the organization will be achieved. Meanwhile, this study contributes to unique theoretical model shows how knowledge-based organizational support, knowledge sharing and AI technologies, knowledge management dynamic capability, and organizational learning (OL) promotes organizational performance and product release through employee skills, technology innovation, and education level factor. As a result, this study's questionnaire and quantitative analysis methods were used for linear regression, PLS -SEM equation and empirical analysis. In summaries, we acknowledge that this study is not without limitations. First, this study only examines 129 observations for data analysis. In the future, we will collect more survey data to support our research. Secondly, we may examine more variables to test our model and research direction from more dimensions. All in all, this study humbly acknowledges that there are some research areas of the limitations in this study. Nevertheless, we hope that the findings of this research will be of use for Chinese technology companies' constructions and those in others countries for them to better grasp the issues and challenges faced by the industry well.

**Acknowledgements**

We would like to express our gratitude to the Solbridge International School of Business, Woosong university faculty and staff for their valuable insights and participation. We thank all friends, managers, classmates, technical architect and professors. This work was supported in part by a grant from Solbridge International School of Business. Moreover, I hope that my technical paper journals can help more and more people understand Knowledge management, KOS, Knowledge sharing with AI, Organizational Learning, KMDC and organizational performance of firms. Thus, we would also like to extend our appreciation to the developers of SPSS 27 and STATA, Smart-PLS software for providing us with a powerful tool that facilitated our data regression analysis and interpretation. In short, I would also like to thank my committee members for letting my defense be an enjoyable moment and for your brilliant comments and suggestions.

**Funding**

The authors declare that no funding was received for the conduct of this research or the preparation of this manuscript. No financial support was provided by any funding agency, institution, or individual. The research was conducted independently, and there are no conflicts of interest to disclose.

**ORCID**

Cui Jun  https://orcid.org/0009-0002-9693-9145

**Declaration of Competing Interest**

The authors declare that they have no conflict of interests

**Appendixes .**

**Appendix A**. Measurement construct items and Sources.

| Variable | Construct Item | Sources |
|---|---|---|
| Knowledge-based organizational Collaborations and support (KOCS) | KOCS1. Knowledge-Based Organizational support and Collaborations (KOS) have by-passed all the above-mentioned limitations, since all the intangible resources can be shared and combined.<br>KOS2. Knowledge-Based Organizational support and Collaborations has assumed the role of Chinese technology companies' strategic resource not only for companies operating autonomously<br>KOS3. KOCS support and Collaboration platform practices promote knowledge creation, sharing, transferring, and utilization necessary for enhanced innovative performance<br>KOS4. knowledge-based CRM and ERP practices as they positively relate to social capital, as they have implications for the density of work-related communication patterns | Singh, S. K., Mazzucchelli, A., Vessal, S. R., & Solidoro, A. (2021) |
| Knowledge sharing KS with AI Technology | KS1. An organization knowledge has processes for absorbing knowledge sharing using AI GPT or AI technologies from individuals into the Chinese technology companies' organization.<br>KS2. An organization of Chinese technology companies has processes for acquiring AI GPT OR AI knowledge from outside organization such as its customers, sub firms or business partners.<br>KS3. An organization of Chinese technology companies has processes for making AI GPT or AI knowledge accessible to those who need it and Chinese technology companies organizing AI knowledge and Using ChatGPT, Copilot or Wenxin AI Model to generate some knowledges. | Kokkaew, N., Peansupap, V., & Jokkaw, N. (2022), Bouschery, S. G., Blazevic, V., & Piller, F. T. (2023), Chowdhury, S., Budhwar, P., Dey, P. K., Joel-Edgar, S., & Abadie, A. (2022) |
| Continue Learning and Training of Chinese technology companies (OL) | OL1. An organization has learned or acquired a lot of new and relevant knowledge AND Organizational members have acquired some critical capacities and continue learning and training new technology skills and courses.<br>OL2. The organization's performance has been influenced by new knowledge learning and employee knowledge training; it has acquired in different departments of Chinese technology companies. | Kokkaew, N., Peansupap, V., & Jokkaw, N. (2022), Rostini, D., Syam, R. Z. A., & Achmad, W. (2022) |
| Knowledge Management Dynamic Capabilities (KMDC) | KMDC1. Improved existing knowledge (codified or not) relevant to issues or problems: concepts; processes/routines; technologies; and products/services (patents, code framework, databases)<br>KMDC2. Employees in different business departments can also share knowledge bases, knowledge articles, and technical information among various departments. | Parent, R., Roy, M., & St-Jacques, D. (2007), Tajpour, M., Hosseini, E., Mohammadi, M., & Bahman-Zangi, B |

| Staff Skills, Technology Innovation, Education Level (SS, TI, EL) | SS. Staff skills include employees' coding skills, communication skills and foreign language skills. The skills of these employees can also ultimately impact the performance of organization in Chinese technology companies.<br>TI. Technological innovation includes innovation in enterprise platforms and software innovation, code framework innovation in enterprise platform's software architecture and innovation in technical code algorithms, etc.<br>EL. The education level of employees also determines the results of organizational performance in Chinese technology companies. | Lynn, G. S., Skov, R. B., & Abel, K. D. (11299). |
|---|---|---|
| Organizational performance of Chinese technology companies (OP). | OP1. An organization puts an emphasis on knowledge share, continue learning and training, and it puts an emphasis on firms' profits, revenue and puts an emphasis on organizational product's release satisfaction, Chinese technology companies' product delivery progress and results.<br>OP2. An organization encourages employees to be innovative in construction methods, knowledge sharing with AI technology, knowledge-based Collaborations and support, knowledge management dynamic capability, and continue learning and training of Chinese technology companies. | Kokkaew, N., Peansupap, V., & Jokkaw, N. (2022), Alavi, M., & Leidner, D. E. (2001), Kitsios, F., & Kamariotou, M. (2021) |